\documentclass{article}

\newcommand{\SSM}{Supersymmetric Standard Model}

\newcommand{\gto}{$SU(2)  \times U(1)$}

\newcommand{\bt}{\begin{tabular}{c}}
\newcommand{\et}{\end{tabular}}

\newcommand{\eb}{\ee\be } 
 
\newcommand{\bmat}{\lt ( \begin{array} }
\newcommand{\emat}{  \end{array} \rt )}

\newcommand{\oQ}{{\ov Q}}

\newcommand{\oq}{{\ov \q}}

\newcommand{\ED}{\end{document}}

\newcommand{\oy}{{\ov \y}}

\newcommand{\oC}{{\ov C}}

\newcommand{\oF}{{\ov F}}
\newcommand{\A}{{\ov A}}

\renewcommand{\a}{\alpha}	
\renewcommand{\b}{\beta}
\newcommand{\g}{\gamma}
\renewcommand{\d}{\delta}
\newcommand{\e}{\epsilon}

\newcommand{\q}{\theta}

\newcommand{\m}{\mu}

\newcommand{\y}{\psi}
\newcommand{\w}{\omega}
\newcommand{\G}{\Gamma}
\newcommand{\D}{\Delta}

\newcommand{\la}{\label}

\newcommand{\ds}{\documentstyle}	
\newcommand{\fr}{\frac}

\newcommand{\pa}{\partial}
\newcommand{\ov}{\overline}
\newcommand{\be}{\begin{equation}}
\newcommand{\ee}{\end{equation}}
\newcommand{\ba}{\begin{array}} 
\newcommand{\ea}{\end{array}}
\newcommand{\bea}{\begin{eqnarray}}
\newcommand{\eea}{\end{eqnarray}}

\newcommand{\Ra}{\Rightarrow}

\newcommand{\Lra}{\Leftrightarrow}
\newcommand{\lt}{\left}
\newcommand{\rt}{\right}

\newcommand{\ben}{\begin{enumerate}}
\newcommand{\een}{\end{enumerate}}
\newcommand{\bitem}{\begin{itemize}}
\newcommand{\eitem}{\end{itemize}}

\begin{document}

\
\vspace{.4in}
\center{\Large CYBERSUSY \\
\Large A new mechanism
 for 
supersymmetry breaking \\
in models like 
the 
\SSM \ (SSM): 

\vspace{.3cm}

{\large John A.  Dixon,
Dixon Law Firm, \\ 1020-833 4th Ave. S.W.,Calgary, Alberta,  Canada T2P 3T5 }
\pagestyle{plain}

\abstract{\normalsize
 The SUSY breaking in Cybersusy is proportional to the VEV that breaks the gauge symmetry \gto \ down to U(1), and it is rather specific to models like the SSM. Assuming full breaking, as explained below, for the leptons, Cybersusy predicts a spectrum of SUSY breaking that is in accord with experimental results so far. In particular, for the choice of parameters below, Cybersusy predicts that the lowest mass superpartner for the charged leptons is a charged vector boson  lepton (the {\bf Velectron}), which has a  mass of 316 Gev . The {\bf Selectron} has a mass of 771 Gev for that choice of parameters. The theory also leads to a zero cosmological constant after SUSY breaking. The mechanism generates equations that restrict models like the SSM.  This version of this paper incorporates recent results and changes discovered subsequent to the talk.}

\section{\large    Composite Superfields when  Auxiliaries are Integrated 
}
\large

Cybersusy arises from a study of the BRS cohomology of D=3+1,  N=1  quantum field theories with chiral and gauged supersymmetry, like the  SSM.  Integration of the auxiliary fields  gives rise to a non-linear realization of supersymmetry, embodied in the nilpotent anticommuting BRS operator $\d_{\rm BRS}$. 
Certain composite expressions, made from the component fields and the Zinn sources,    behave  almost as though they were superfields.  The non-linearity implies `Cybersusy Constraints' that must be satisfied  to form   composite `pseudosuperfields'.  
When the vacuum expectation value that breaks gauge symmetry becomes non-zero, one finds that  these composite  pseudosuperfields give rise to a new `anomalous' realization of supersymmetry.  Mapping the composite superfields onto new elementary effective superfields yields a new action with broken supersymmetry.  	

It turns out that the SSM is an excellent model in which to find solutions to the Cybersusy Constraints. The new realization that arises, when the VEV appears, gives rise to a natural mechanism for SUSY breaking.  And those solutions, and that SUSY breaking, look very much like the particles we observe. Here we only consider the leptons, but it does appear that the mechanism extends to most or all  other particles.  Cybersusy is an acronym standing for `{\bf C}ohomologicall{\bf Y} {\bf B}roken
 {\bf E}ffective {\bf R}etro {\bf SU}per{\bf SY}mmetry'. 
The word {\bf R}etro refers to the fact that Cybersusy leads us back to composite particles like baryons which were the ultimate origin of supersymmetry, through duality, the string and the superstring.

\section{\large   BRS operator for   chiral matter in a general theory } 

Our considerations are restricted to the the chiral matter here. However, it appears that nothing essential changes when the gauge theory is introduced. The action is
${\cal A}_{\rm SSM }  =
\int d^4 x \; d^4 \q
{\hat A}^p {\hat {\ov A}}_p+ 
\int d^4 x \; d^2 \q
 \lt \{
\fr{1}{3}  g_{pqr}  
{\widehat A}_{}^{p} 
{\widehat A}_{}^{q} 
{\widehat A}_{}^{r} 
-  m^2 g_r {\widehat A}_{}^{r} 
\rt \}
+ *
$. The SSM  develops a VEV $< A^p>$ satisfying $< A^p> = m v^p $ and $ g_{pqr} v^q v^r - g_r=0
$ which breaks \gto \ down to U(1). 
We add Zinn Justin terms 
$Y_i^{\a} \d_{\rm WZ} \y^i_{\a} 
=
Y_i^{\a} \lt ( \pa_{\a \dot \b} A^i 
\oC^{\dot \b} + F^i C_{\a} \rt )
$ and $ 
\G_i \d_{\rm WZ} A^i= 
\G_i C^{\a} \y^i_{\a} 
$
to the Lagrangian, place this in a Feynman path integral, observe the supersymmetry,  
 integrate $F^r$, and perform the usual steps to get a nilpotent $\d_{\rm BRS}$ which satisfies $
\d^2_{\rm BRS}= C^{\a} \oC^{\dot \b} \pa _{\a \dot \b} 
\approx 0
$.  Here is the operator:  
\[
\d_{\rm BRS} =
\int d^4 x\;
  \y^{i}_{  \b} {C}^{  \b} 
\fr{\d  }{\d A^i} 
+
\int d^4 x\;
\lt \{
\pa_{ \a \dot \b }  A^{i} {\ov C}^{\dot \b}  
+
C_{\a}  G^i 
\rt \}
\fr{\d  }{\d \y_{\a}^i} 
+
\int d^4 x\;
\]
\[
\lt \{
 - \fr{1}{2} \pa_{ \a \dot \b  }       \pa^{ \a \dot \b  }        {\ov  A}_{i} 
-
\pa_{ \a \dot \b } Y_{i}^{ \a}    {\ov C}^{\dot \b}   
   + g_{ijk} \lt [
   2  A^{j}   
G^k - \y^{j \a} 
\y^{k}_{ \a}  
 \rt ]
 +
2 m g_{ijk} v^j  G^{k}     
\rt \}
 \fr{\d}{\d \G_i } 
\]
\be
+
\int d^4 x\;
\lt (
-
  \pa^{\a \dot \b  }   
{\ov \y}_{i   \dot \b}
 +
2 g_{ijk}  \y^{j \a} A^k    
 +
2 m g_{ijk}  \y^{j \a} v^k    
-
\G_i  
 {C}^{  \a}
\rt )
 \fr{\d  }{\d Y_{i}^{ \a}} 
+ *
\la{deltageneral}
\ee
The composite field $G^i$ is 
$G^i =  - \lt (    
 {\ov g}^{ijk} {\ov A}_j  {\ov A}_k  +
2
m {\ov g}^{ijk} {\ov A}_j  {\ov v}_k  +
{\ov Y}^{i \dot \b} {\ov C}_{\dot \b } 
\rt )
$.

\section{\large  Fundamental  Superfields  with integrated auxiliaries}

Certain combinations of fields,  sources and $\q, \ov \q$ act like superfields. We will call them {\bf Fundamental   Superfields}. The first is the  {\bf Fundamental chiral superfield} $
{\hat A}_{\rm Fund}^{i}(x) = A^i(y) +
\q^{\a} 
\y^i_{\a } (y) 
+ \fr{1}{2} \q^{\g} \q_{\g} 
G^i(x)$ where the translated spacetime variable is $y_{\a \dot \b}  = x_{\a \dot \b} + \fr{1}{2} \q_{\a } \ov \q_{\dot\b} 
$. The transformation induced by $\d_{\rm BRS} $ is summarized by the following equation:
$
\d_{\rm BRS}  {\hat A}_{\rm Fund}^{i}(x)=   \d_{\rm SS} {\hat A}_{\rm Fund}^{i}(x) $ where the superspace operator is $\d_{\rm SS} =
C^{\a} Q_{\a} 
+ \ov C^{\dot \a} \ov Q_{\dot \a} 
$.  This relation means that the effect of  $\d_{\rm BRS}$ on this particular combination is the same as the effect of the superspace operator $\d_{\rm SS}$.  The supertranslations are: 
$
 Q_{\a} = \fr{\pa}{\pa \q^{\a}} -  \fr{1}{2} \pa_{\a \dot \b} \oq^{\dot \b} $ and $  \oQ_{\dot \a} = \fr{\pa}{\pa \oq^{\dot \a}} -  \fr{1}{2} \pa_{\b \dot \a} \q^{\b}
$.
 {\bf Next is the surprise. There is a  new kind of Superfield which is not present in the usual treatment!} {\bf It is the Fundamental  chiral dotted spinor
 superfield: }
$
{\hat {\ov \y}}_{{\rm Fund}\; i \dot \a}(x) = \ov \y_{i \dot \a }(y) 
+
\q^{\b} 
\lt [
\pa_{\b \dot \a} \A_i(y) 
+ \ov C_{\dot \a} Y_{i  \b}(y) \rt ]
- \fr{1}{2} \q^{\g} \q_{\g} 
\G_{i}(x)  
\ov C_{\dot \a} 
$.   Its transformation under the action of  
$ \d_{\rm BRS}$ is: 
$
\d_{\rm BRS}  {\hat {\oy}}_{{\rm Fund}\;i \dot \a}(x)=   \d_{\rm SS} {\hat \oy}_{{\rm Fund }\;i \dot \a }(x)  -
 g_{ijk} 
  {\hat A}_{\rm Fund}^j {\hat A}_{\rm Fund}^k 
\ov C_{\dot \a} 
-
2m g_{ijk} 
  v^j {\hat A}_{\rm Fund}^k 
\ov C_{\dot \a} 
$. It  behaves as a chiral superfield if and only if the theory is free and massless,  which happens if and only if
$g_{ijk} =m^2 g_{i} =0$.  Its nonlinear transformation suggests that we form composite superfields as follows.

\section{\large  Composite Superfields   in general theory}
Consider  the composite expression
$
{\hat \w}_{{\rm Comp}\;\dot \a} = f^{i}_{j} {\hat {\ov \y}}_{{\rm Fund}\; i \dot \a}
\lt ( m v^j + {\hat A}_{\rm Fund}^{j}\rt )
$. The constraint equations are
$f^{i}_{(j}   g_{kl)i}  =0
$ and if they are satisfied, then we get $
\d_{\rm BRS}  {\hat \w}_{{\rm Comp}\; \dot \a} = \lt ( 
\d_{\rm SS}  + \d_{\rm GSB}  \rt ){\hat \w}_{{\rm Comp}\; \dot \a}
$, where the new variation is 
$
\d_{\rm GSB}\; {\hat \w}_{{\rm Comp}\; \dot \a}=  m^2 f^{i}_{j}  g_i  {\hat A}_{\rm Fund}^{j} 
\ov C_{\dot \a} 
$. $ {\rm GSB} $ stands for gauge symmetry breaking. If $m^2 g_i =0 $, then  ${\hat \w}_{{\rm Comp}\; \dot \a}$ behaves as a superfield ($
\d_{\rm BRS}  = 
\d_{\rm SS}  
$) 
and if $m^2 g_i \neq 0 $, then  ${\hat \w}_{{\rm Comp}\; \dot \a}$ has a new term in the algebra, namely 
$\d_{\rm BRS}  =  
\d_{\rm SS}  + \d_{\rm GSB}  
$.

\section{\large  Solutions of Cybersusy Constraints  for  SSM  }

 The SSM superpotential has the following form 
in terms of the usual Quark, Lepton, and Higgs  doublet and  singlet multiplets:
\be
P_{{\rm SSM} }
=
g \e_{ij} H^i K^j J
+
p_{p q} \e_{ij} L^{p i} H^j P^{ q} 
+
r_{p q} \e_{ij} L^{p i} K^j R^{ q}
\eb
+
t_{p  q} \e_{ij} Q^{c p i} K^j T_c^{ q}
+
b_{p  q} \e_{ij} Q^{c p i} H^j B_c^{ q}
- m^2 g_{J } J \ee

The term $- m^2 g_{J } J$ yields VEVs:
$
<H^i> = m h^;\; <K^i> = m k^i
$.  These break  $SU(2) \times U(1) \Ra U(1)$, but there is no  spontaneous breaking of SUSY, because the auxiliary fields have zero VEV:
$
< F^p> = < D^a> =0
\Lra 
{\rm Zero\; Vacuum \; Energy}
\Lra 
{\rm Zero\; Cosmological\; Constant}
$.
Now we look at the {\bf SSM in detail} to find solutions of the constraint equations$
f^{i}_{(j}   g_{kl)i}  =0
$.  The SSM (and related models) provide {\bf surprising} examples of these. 
Observe that
$
(f^{i}_{(j}   g_{kl)i}  =0 )
\Lra ({\cal L}  P_{\rm Cubic} = 0)
$ where ${\cal L} $ is a Lie algebra generator:
${\cal L} = 
  f^{m}_{n}    {  A}^{n} \fr{\pa}{\pa {  A}^{m}}
$ and $ P_{\rm Cubic} =  g_{ijk} { A}^{i}
{  A}^{j} {  A}^{k} $.

It is easiest at first to look for generators with nonzero baryon or lepton number, to avoid dealing with the gauge theory.

\section{\large  Generators for Charged Leptons  in the SSM:}

The operators   
$
{\cal L}^+_{p}=p_{pq} P^q
\fr{\pa}{\pa J} + 
g   K^j \fr{\pa}{\pa L^{pj}}$, $
{\cal L}^-_{p}= p_{qp} \e_{ij} H^i L^{qj} 
\fr{\pa}{\pa J} + 
g \e_{ij} H^i K^j \fr{\pa}{\pa P^{p}} 
$ both satisfy ${\cal L} P_{\rm Cubic}=0$ for the SSM.  Each invariant  
Lie algebra operator  yields a  chiral dotspinor superfield.  For example 
${\cal L}^+_{p}=p_{pq} P^q
\fr{\pa}{\pa J} + 
g   K^j \fr{\pa}{\pa L^{pj}} 
 \Lra 
{\hat {\w}}^+_{{\rm Comp}\; p, \dot \a}=p_{pq} {\hat P}^q
{\hat {\ov \y}}_{J \dot \a} + 
g   {\hat K}^j {\hat{\ov \y}}_{L pj \dot \a}
$. The next step is to map these composite fields 
${\hat {\w}}^+_{{\rm Comp}\; p, \dot \a}$ onto elementary effective fields, 
${\widehat \w}_{R p\dot \a} $, and to deduce the algebra of the effective fields from the algebra of the  composite fields: 
\be
\begin{array}{|cccc|}  
\hline
\multicolumn{4}{|c|}{\rm   Cybersusy \; Effective  \; Superfields \; from \; the \; SSM \;for \;the\; Charged\; Leptons}
\\
\hline
{\rm Effective }& {\rm Composite }   &
  {\rm Y} 
&
{\rm L}
\\
\hline
{\widehat A}_{L}^{p} 
 & {\widehat L}^{ip} ( m h_i + {\widehat H}_i )&
  -2
& +1
\\
\hline
{\widehat A}_{R}^{p}
 & {\widehat P}^{p}  &
  +2
& -1
\\
\hline
{\widehat \w}_{L p \dot \a}
 & 
\begin{array}{c}
p_{qp}( m  h^j + {\widehat H}^j ) {\hat L}^{q}_j 
{\hat {\ov \y}}_{J \dot \a} 
+ 
g   ( m  h^j + {\widehat H}^j )
\\
( m  k_j + {\widehat K}_j )
{\hat{\ov \y}}_{P p \dot \a}
\end{array}
 &
  -2
& +1
\\
\hline
{\widehat \w}_{R p\dot \a} 
 & 
\begin{array}{c}
p_{pq} {\hat P}^q
{\hat {\ov \y}}_{J \dot \a}  
+ 
g   ( m  k^j + {\widehat K}^j )
 {\hat{\ov \y}}_{L p j \dot \a}\end{array}
 &
  +2
& -1
\\
\hline
\end{array}
\la{quartet2}
\ee
In the above, p=1,2,3 is a flavour index. Y is electric hypercharge and L is lepton number. These superfields are singlets under SU(2) and SU(3):

The operator $\d_{\rm BRS}  =  
\d_{\rm SS}  + \d_{\rm GSB} $ acting on the composite fields implies a corresponding  algebra for the effective fields, namely $
\d_{\rm Cybersusy}\equiv \d_{\rm CS}= \d_{\rm SS}+\d_{\rm Mix}
$.
The new variations are
$
 \d_{\rm Mix}  {\widehat \w}_{L p\dot \a}=p_{qp} {\widehat A}_{L }^{q}
\ov C_{\dot \a}
; 
 \d_{\rm Mix}  {\widehat \w}_{R p\dot \a} =  p_{pq}  
{\widehat A}_{R }^{q} \ov C_{\dot \a}
; \d_{\rm Mix}  {\widehat A}_{L}^{q} =  0
; \d_{\rm Mix}  {\widehat A}_{R }^{q} =  0
$.
\section{\large  Effective Fields and Action  for Charged Leptons}

Now we look for  a new action expressed in terms of the above effective fields.   We want this to be invariant under the new transformation 
$\d_{\rm CS}= \d_{\rm SS}+\d_{\rm MIX}
$. So we start with:
$
\d_{\rm SS}{\cal A}_{\rm   WZ}=0
$ and then look for ${\cal A}_{\rm  Compensator}$ to satisfy:
$\d_{\rm MIX}
{\cal A}_{\rm   WZ}
+\d_{\rm SS}  {\cal A}_{\rm  Compensator}=0
$.  First we need a {\bf Kinetic Compensator}  ${\cal A}_{\rm KCL} =  {\cal A}_{\rm KCL1} + {\cal A}_{\rm KCL2}$
The action  so far takes the form:
\[
\begin{array}{|cc|}  
\hline
\multicolumn{2}{|c|}{ {\cal A}_{\rm WZ}  }
\\
\hline
\hline
{\rm Name}& {\rm Action} 
\\
{\cal A}_{\rm  ScalarL}
& \fr{1}{4}\int d^4 x \; d^4 \q
{\widehat A}_{L }^{p} 
{\widehat {\ov A}}_{L p} \\
{\cal A}_{\rm  ScalarR} & L \Ra R
\\
{\cal A}_{\rm Scalar\;  Mass}  &
\fr{1}{2}\int d^4 x \; d^2 \q
  g_{pq} m 
{\widehat A}_{L}^{p} 
{\widehat A}_{R}^{q} + *
\\
{\cal A}_{\rm  DotspinorL}
& -\fr{1}{4}\int d^4 x \; d^4 \q
{\widehat {\ov \w}}_{L  \a }^{p} 
\pa^{\a \dot \b}
{\widehat \w}_{L p\dot \b} 
\\
{\cal A}_{\rm  DotspinorR}
& L \Ra R\\
{\cal A}_{\rm Dotspinor\;  Mass}  
&\fr{1}{2}\int d^4 x \; d^2 \ov \q
\ov d_{pq} m^2 {\widehat {\ov \w}}_{L  \a }^{p} 
{\widehat {\ov \w}}_{R  }^{ \a q}  
+ * 
\\
\hline
\end{array}
\begin{array}{|cc|}  
\hline
\multicolumn{2}{|c|}{{\cal A}_{\rm KCL} }
\\
\hline
\hline
{\rm Name}& {\rm Action} 
\\
{\cal A}_{\rm KCL1}
&\int d^4 x \;
\lt ( 
{\ov p}^{q  p }
\ov \y^{}_{{L}     \dot \a q}
\D 
\w^{  \dot \a}_{{L}   
  p }
\rt.
\\
&\lt.
+
{\ov p}^{q  p }
\ov F^{}_{{L} q  }
\pa_{\a \dot \a}
W^{  \a \dot \a}_{{L}   
 p  }
\rt.
\\
&\lt.
+ 
 {p}_{q  p }
 \y^{q  }_{{L}    \a}
\D 
\ov \w^{   p  \a}_{{L}   
 }
\rt.
\\
&\lt.
+
 {p}_{q  p }
 F^{ q }_{{L} }
\pa_{\a \dot \a} 
\ov W^{  p  \a \dot \a}_{{L}   
 }
\rt )
\\
{\cal A}_{\rm KCL2}
&\int d^4 x \;
 {p}_{q  p }
{\ov p}^{s  p }
\lt (
  A^{  q  }_{{L} }
\D 
\A^{ }_{{L}   s }
\rt.
\\
&\lt.
+
 F^{ q   }_{{L} }
 \oF^{ }_{{L}  s  }
\rt )
\\
\hline
\end{array}
\la{effaction2}
\]
Adding the ${\cal A}_{\rm KCR}$ action yields a new action
$
{\cal A}_{\rm Cybersusy}={\cal A}_{\rm WZ}+ {\cal A}_{\rm KCL}+ {\cal A}_{\rm KCR}
$. {\bf Next} we look for a  {\bf Mass  Compensator}  ${\cal A}_{\rm MC1} $.   It needs to satisfy
$
  \d_{\rm MIX}{\cal A}_{\rm  Dotspinor\; Mass}+  
\d_{\rm SS}{\cal A}_{\rm  MC1} =0
\la{firststepmass}
$. {\bf It is easy to show that no such  local Mass  Compensator
${\cal A}_{\rm MC1}$ can exist}, because  
$
\d_{\rm MIX} {\cal A}_{\rm Dotspinor\;  Mass}  
=
m^2 {\cal A}_{\rm   Anomaly} 
\in  {\rm  Cohomology\; of }
\; \d_{\rm SS}
$. But $\d_{\rm MIX}=0 $ implies that there is  no gauge symmetry breaking, and ${\cal A}_{\rm Dotspinor\;  Mass}  =0$    implies that there is a massless charged lepton supermultiplet. The only sensible choice is
that $m^2 {\cal A}_{\rm   Anomaly} 
\neq 0 $.
The anomaly comes from the algebra, the action, and physical reasoning, {\bf not from a loop diagram}. This    uniquely defined  action yields SUSY breaking proportional to gauge symmetry breaking.   

After this talk was given,  the author realized that the $\d_{\rm Mix}$ part of the algebra disappears for the left sector if one chooses the left composite operator to be 
${\widehat \w}_{L p \dot \a}
\approx
p_{qp}( m  h^j + {\widehat H}^j ) {\hat L}^{q}_j 
{\hat {\ov \y}}_{J \dot \a} 
+ 
\lt \{
g   ( m  h^j + {\widehat H}^j )
( m  k_j + {\widehat K}_j )
- m^2 g_J
\rt \}
{\hat{\ov \y}}_{P p \dot \a}
$ in place of 
 (\ref{quartet2}).  No such possibility arises for the right sector. So Cybersusy still breaks SUSY after this change,  but a reasonable spectrum may require a modification of the model.
 
\section{\large   Broken SUSY Spectrum for one flavour of Charged Leptons}

We now describe the mass spectrum for the simplest case of one flavour, assuming both left and right breaking, as it arises from table  (\ref{quartet2}). G, P and D are positive parameters. Define 
$
X = \fr{p_{\m} p^{\m}}{m^2}
$.  The fermionic lepton masses are the negative solutions of
$ P_{\rm Quintic \; Fermi}(X) =  X 
 \lt \{ X^2 (  1-P) -  D \rt \}^2 
+      G  \lt \{ X^2    -  D \rt \}^2 =0
$. The bosonic lepton masses are the negative  solutions of
$ P_{\rm Quadratic \; Bose }(X) = X^2  - D=0 
\la{wterjrtfuj3}
$ and
$P_{\rm Quartic \; Bose }(X)
 = X^2  
\lt (  X  (1-P)^2  +  G  
\rt )^2 
- \lt ( X(1-P^2)+G \rt )^2  D 
=0
$.  The following choice of parameters is interesting: $
P= 1-10^{-5.8} , G = 2.5*10^{-7}, D = 10^{10}$.  It yields two very heavy fermionic leptons with masses 
  8992  and 8834 Gev,  plus the light  
{\bf Electron} with mass  $. 5 *10^{-3}$   Gev.  Then there is one very heavy  scalar  boson lepton with mass 
  355,000 Gev, and a much lighter scalar  boson lepton, the 
{\bf Selectron } with mass 771  Gev.  The lightest superpartner for this choice of parameters is 
the  vector boson lepton, the  
 {\bf  Velectron}, with mass  316   {\rm Gev}. 

\vspace{.2cm}

\begin{center}
{\bf {\Large Reference}}

\end{center}

More complete references can be found in J. A. Dixon,  Cybersusy I,  arXiv: 0808-0811 hep-th, Aug 6, 2008 and  Cybersusy II-V, also on arXiv.  These papers require some revision to incorporate the modification to equation 
 (\ref{quartet2}) discussed above.

\end{document}